\documentclass[aps,prl,twocolumn,citeautoscript,superscriptaddress]{revtex4-2}
\usepackage{amsmath}
\usepackage{amssymb}
\usepackage{graphicx,bm}
\usepackage{verbatim}
\usepackage{epstopdf}
\usepackage{dcolumn}
\usepackage{floatrow}
\usepackage{float}
\usepackage{booktabs}
\usepackage[utf8]{inputenc}
\usepackage{array}
\usepackage{makecell}
\usepackage{booktabs}
\usepackage{multirow}
\usepackage{dcolumn}
\usepackage{footnote}

\usepackage{lipsum}
\usepackage{hyperref}
\hypersetup{
	colorlinks=true,
	linkcolor=magenta,
	filecolor=violet,     
	urlcolor=blue,
	citecolor=red
}

\begin{document}
	
\title{Design and characterization of a recoil ion momentum spectrometer for investigating molecular fragmentation dynamics upon MeV energy ion impact ionization}

\author{Avijit Duley, Rohit Tyagi, Sandeep B. Bari, and A. H. Kelkar}
\email{akelkar@iitk.ac.in}
\affiliation{Department of Physics, Indian Institute of Technology Kanpur, Kanpur - 208016, India}

\begin{abstract}
We present the development and performance of a newly built recoil ion momentum spectrometer to study the fragmentation dynamics of ionized molecules. The spectrometer is based on the two-stage Wiley-McLaren geometry and satisfies both time and velocity focusing conditions. An electrostatic lens has been introduced in the drift region to achieve velocity imaging and higher angular collection. The spectrometer is equipped with a 2D position-sensitive detector with multi-hit coincidence electronics. Ionic fragments with kinetic energy $\sim$ 8 eV can be detected with 4$\pi$ collection. The overall performance of the spectrometer has been tested by carrying out three-dimensional ion imaging measurements for diatomic (N$_2$) and polyatomic (CH$_2$Cl$_2$) molecules under the impact of 1 MeV proton. Three-dimensional momentum and kinetic energy release distributions were derived from the measured position and time-of-flight spectra. The observed features of the various fragmentation channels as well as the measured kinetic energy release distributions are in complete agreement with the available data.
\end{abstract}

\maketitle
\section{Introduction}
The study of dynamics of molecular fragmentation has evolved as an important field of research in atomic collision physics. Numerous theoretical and experimental reports have investigated the dissociation dynamics of simple and complex molecules following photon and particle-induced ionization. Unlike ionization of atomic systems, molecular ionization and subsequent fragmentation is governed by the non-Coulombic interaction potential. A complete \textsl{ab initio} theoretical description for molecular fragmentation becomes rather complex, and high-quality experimental data often guides theoretical advancements. Ullrich and Schmidt-B{\"o}cking \cite{Ullrich_1987, Olson_1987} pioneered the technique of recoil ion momentum spectroscopy to measure individual momenta and kinetic energy of all particles participating in a collision event, which has now become central to the investigation of correlated dynamics in many-body systems. Several modifications of the original spectrometer have been developed over the years. This technique employs the extraction of charged fragments created in a collision event using electric fields (and magnetic fields in some cases). These fragments are further detected by a large area position-sensitive detector coupled with multi-hit coincidence electronics. The time-of-flight and hit positions on the detector are measured for individual fragment ions. The initial momenta and kinetic energy of the fragment ions can be reconstructed using this information, thereby revealing the details of the ionization and fragmentation process.

In this paper, we discuss the design and performance of a newly developed recoil ion momentum spectrometer (RIMS), which can be utilized to study ionization-induced fragmentation dynamics of molecules. The spectrometer has been developed and installed in the Tandetron Accelerator Laboratory at IIT Kanpur, India. In a RIMS, target molecules are made to interact with the projectile ion beam in a well-localized region. During a collision with the energetic ion beam, various charge transfer processes such as ionization, capture, and transfer ionization create recoil ions, which are then extracted and guided towards a 2D position-sensitive microchannel plate detector using a combination of electric fields. The initial momenta (along the three space directions) of each fragment ion are reconstructed from the time-of-flight (TOF) and hit positions on the detector. We have designed a two-stage Wiley-McLaren \cite{Wiley_1955} type spectrometer satisfying both time and space focusing conditions. The spectrometer has been slightly modified by introducing an electrostatic lens in the drift region. This lens produces a non-uniform electric field, which is optimized to obtain velocity imaging as well as higher $4\pi$ collection efficiency. The introduction of a lens to achieve both time and space focusing was first attempted by Lebech \emph{et al} \cite{lebech_2002}. Later, Schmidt \emph{et al} \cite{schmidt_2004} and Bomme \emph{et al} \cite{bomme_2013} also employed a lens to achieve larger collection efficiency. Prumper \emph{et al} \cite{prumper_2007} also used a lens for higher $4\pi$ collection efficiency and proper velocity focusing without satisfying the Wiley-McLaren time focusing condition. Recently, there have been a few reports \cite{laksman_2013,khan_2015,biswas_2021} where the authors have combined both time and space focusing in a two-field Wiley-McLaren type spectrometer with a higher $4\pi$ collection efficiency. 

To study the performance of the spectrometer, we have measured the fragmentation channels and kinetic energy release distributions along with the angular distribution for 1 MeV proton impact ionization of N$_2$ (diatomic) and CH$_2$Cl$_2$ (polyatomic) molecules. 

\section{Spectrometer Design and Experiment}

\subsection{Experimental details}

\begin{figure*}
\includegraphics[width=\textwidth]{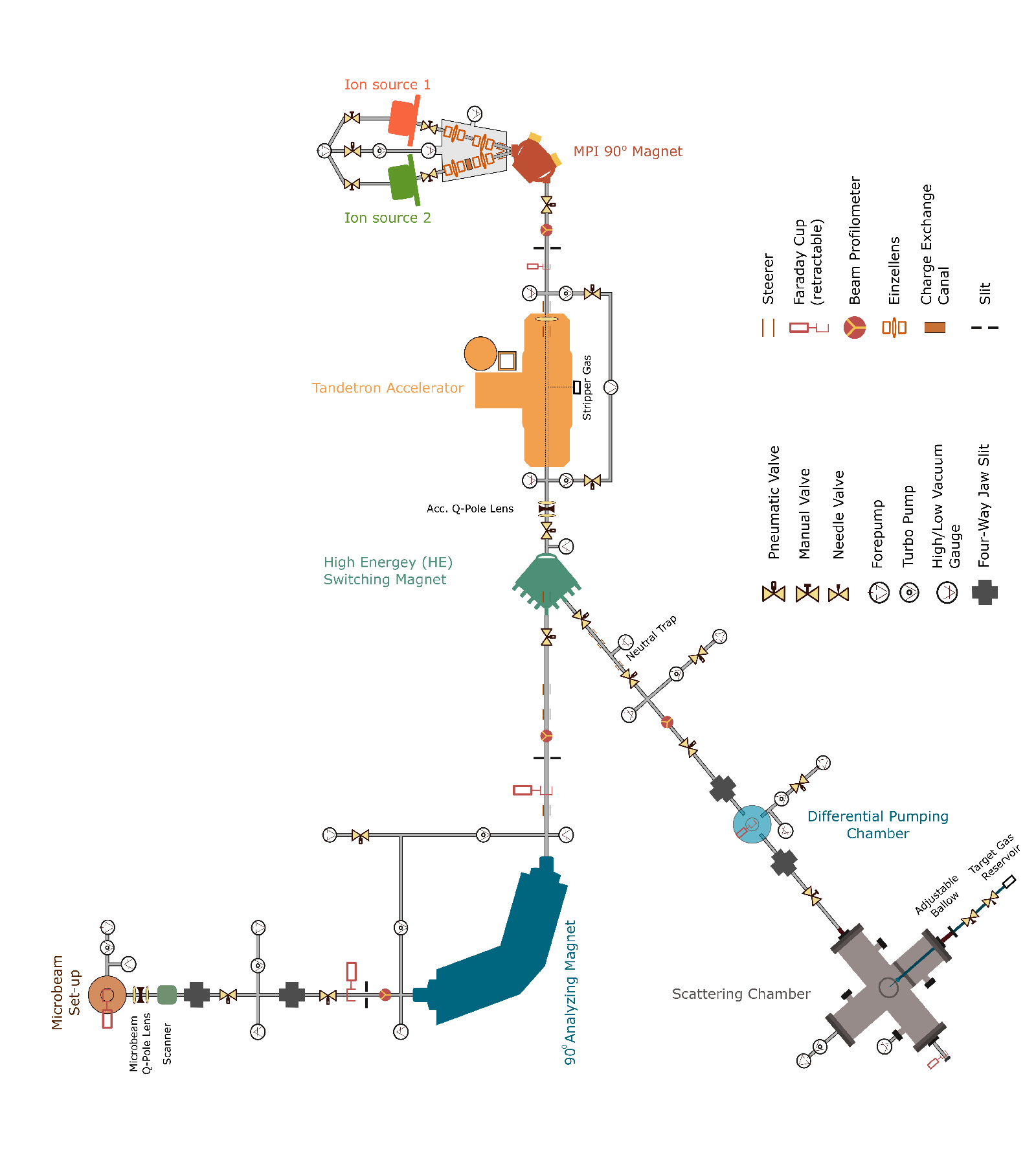}
\caption{Schematic layout of the 1.7 MV (high current) tandetron accelerator facility, IIT Kanpur, India.}
\label{Fig.1}
\end{figure*}

The experiments were performed at the 20$^{\circ}$ beam line of the 1.7 MV (high current) tandetron accelerator \cite{hvee} facility, IIT Kanpur, India. A schematic layout of the accelerator facility is shown in figure \ref{Fig.1}. A beam of low energy (30 keV) H$^-$ ions is produced from a duoplasmatron source and injected into the accelerator after mass selection. MeV energy H$^+$ ions are subsequently extracted on the high-energy side of the accelerator. The H$^+$ beam is focused using an electrostatic quadrupole triplet, and a switching magnet steers the beam along the 20$^o$ beam line. While traversing through the beamline, the ion beam may acquire some fraction of neutral H atoms due to charge exchange processes. A neutral beam trap is installed near the end of the beamline to separate the neutrals from the ion beam. The projectile beam is further collimated using a pair of four-way jaw slits, located approximately 1 m apart, before entering the main scattering chamber. The vacuum in the accelerator system was maintained at $\sim$ $\text{1}\times \text{10}^{\text{-7}}$ mbar.

\begin{figure*}
\includegraphics[width=0.8\textwidth]{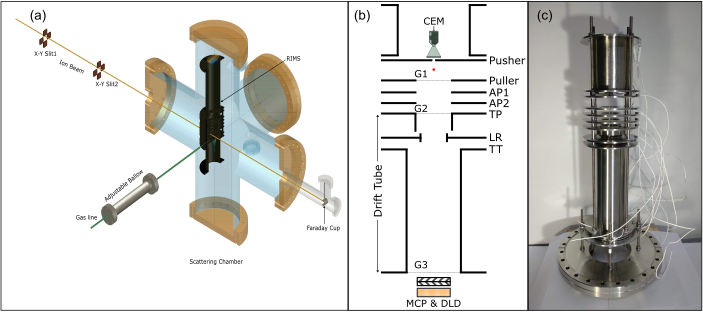}
\caption{(a) Schematic layout of the collision geometry consisting of a collimated projectile ion beam, a target gas, and the RIMS along with a cut-view of the scattering chamber. (b) Schematic diagram of the recoil ion momentum spectrometer. G1, G2, and G3 are the three high transmission metallic grids. LR is the lens ring. The red dot at the center of the extraction region indicates the interaction region. (c) A photograph of the actual spectrometer assembly without the position sensitive detector.}
\label{Fig.2}
\end{figure*}

The main scattering chamber is a six-way cross with CF-250 end ports (see figure \ref{Fig.2}). The scattering chamber is pumped through a single 700 l/sec turbo molecular pump backed by a rotary vane pump. The background pressure in the main scattering chamber is maintained at $\le \text{5}\times \text{10}^{\text{-8}}$ mbar. The ion beam is collected by a semicircular FC kept at a distance of 900 mm from the interaction zone at the end of the scattering chamber. The spectrometer is mounted in the vertical arm of the main scattering chamber perpendicular to the projectile beam direction. The target gas is introduced in the interaction region as an effusive jet using a high aspect ratio ($\sim \text{0.008}$) hypodermic needle kept at ground potential. The needle is connected to a flexible bellow, and its position with respect to the interaction region can be adjusted. The needle is at the center of the extraction region and does not affect the electric field in the extraction region. The projectile ion beam, the target effusive gas jet, and the spectrometer axis are aligned mutually perpendicular. The backing pressure in the gas line is adjusted using a needle valve to maintain a background pressure of $\sim \text{1}\times\text{10}^{\text{-6}}$ mbar in the scattering chamber during the experiments.  


\subsection{Recoil Ion Momentum Spectrometer}

The spectrometer is modeled on the two-stage Wiley-McLaren \cite{Wiley_1955} geometry. It consists of three regions, namely,  an extraction region, an acceleration region, and a drift region. The spectrometer electrodes and supporting rods are made of non-magnetic stainless steel (SS-316). All the electrodes have an outer diameter of 120 mm. A schematic design of the spectrometer is shown in figure \ref{Fig.2}(b). The distance between the pusher and the puller electrodes is 20 mm. These two electrodes form the extraction region. The inner diameter of the puller electrode is 40 mm, while the pusher electrode has a small 2 mm hole in the center. The opening in the pusher electrode is deliberately kept small to restrict the background electron count rate. The electrodes are 3 mm thick. The region between the puller electrode and the drift tube forms the acceleration region having a length of 38 mm. Two electrodes, AP1 and AP2 (OD = 120 mm and ID = 40 mm) are placed in the acceleration region, equidistant apart, to maintain a uniform acceleration field. The field-free drift region is divided into two parts separated by a lens ring (LR). The lens ring is placed at a distance of 25 mm from the drift tube entrance and acts as an Einzel lens to focus the recoil ions on the ion detector.  
The total length of the drift region is 245 mm. The puller electrode and the drift tube entrance and exit apertures are covered with nickel mesh (G1 - G3, transmission = 90\%) to maintain field uniformity in the interaction region and drift region. The drift region is followed by a position-sensitive detector (PSD) assembly. The PSD consists of two microchannel plates (MCPs) (in chevron configuration) and a delay line anode (DLD-40) \cite{Jagutzki_2002}. The MCPs have an active diameter of 40 mm.

An electric field of 145 V/cm was used in the extraction region. The field in the acceleration region was set at 260 V/cm. The interaction zone remains at 0 V since a symmetric voltage ($\pm$ 145 V) has been used in the extraction region. The two parts of the drift tube (TP and TT) are shorted internally and are kept at -940 V. Two distinct operational conditions are defined depending on the voltage applied to the lens ring. 1) Lens OFF, where the lens ring is at the same potential (-940 V) as the drift tube, and 2) lens ON when there is a different voltage on the lens ring. In the present experiment, the lens ring was kept at -500 V in the lens ON condition. The voltages on all the electrodes can be controlled separately. A channel electron multiplier (CEM, Sjuts KBL RS10) is mounted on the backside of the pusher electrode to detect the electrons ejected during a collision event. 

\subsection{Simulation Results}
\begin{figure}
\includegraphics[width=0.9\textwidth]{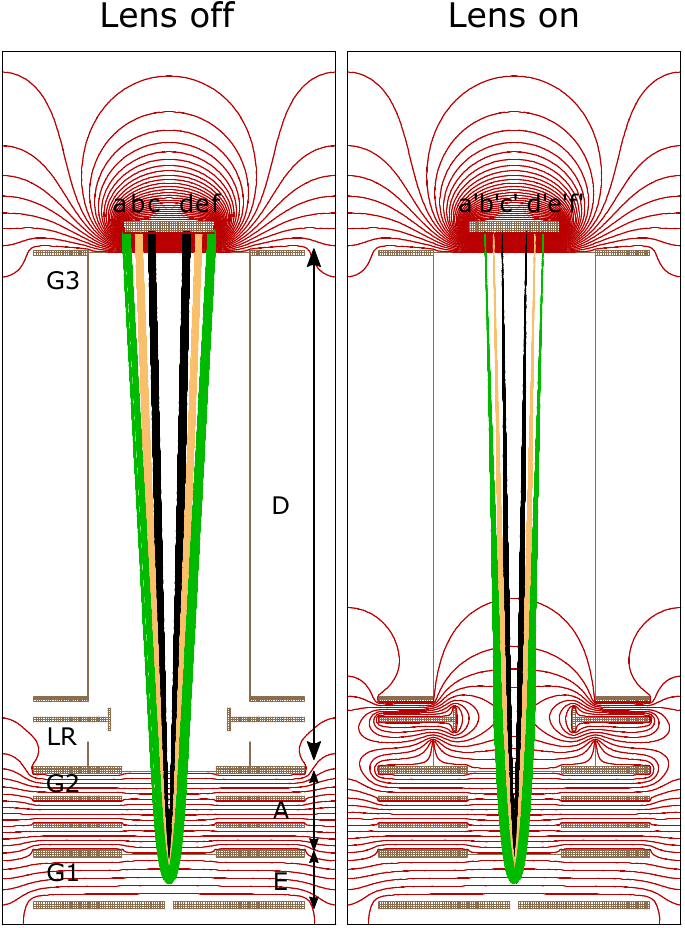}
\caption{Ion trajectory simulation of the recoil ion momentum spectrometer (RIMS) using SIMION with lens ON and OFF conditions. The trajectories for ions with three different energies are shown in black (0.5 eV), yellow (1.5 eV), and green (3 eV). The ion-hit positions on the detector are indicated by a,b,c,d,e for the lens OFF condition, whereas a$'$,b$'$,c$'$,d$'$,e$'$ denote the same for lens ON condition.}
\label{Fig.3}
\end{figure}

\begin{figure*}
\raggedright
\includegraphics[width=\textwidth]{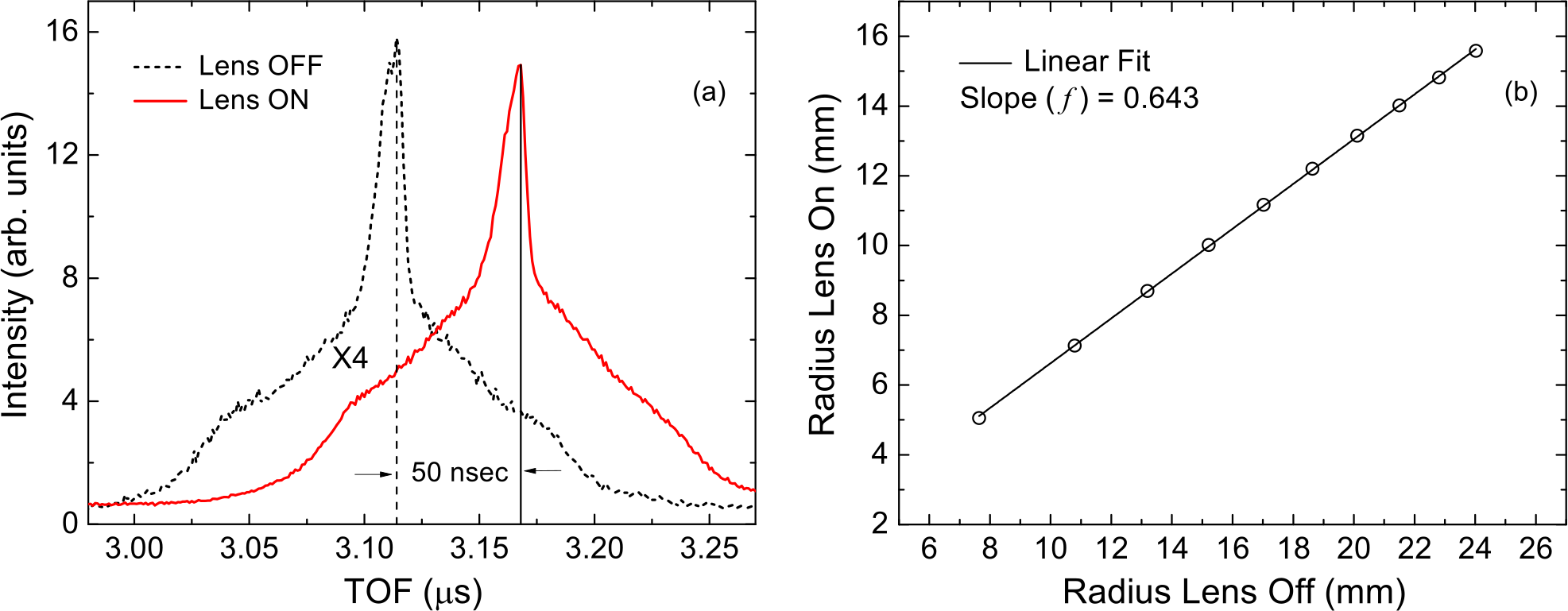}
\caption{(a) Measured difference in TOF for the lens ON and OFF condition. Vertical lines depict the peak position of the two spectra which corresponds to the particles with zero initial momentum along the spectrometer axis. (b) Simulated hit radius of ions with lens ON and OFF condition shown by open circles. The solid line is the linear fit to the simulated data. The slope of this curve gives the value of the "bending coefficient" ($f$).}
\label{Fig.4}
\end{figure*}

A detailed simulation was performed using the ion optics simulation program SIMION 8.0 \cite{simion}. In the present design, we have tried to achieve three-dimensional focusing of the ions by introducing simultaneous time and velocity focusing. Optimization of these focusing conditions was done by considering a uniform distribution of ions within a sphere of 2 mm diameter having an isotropic kinetic energy distribution in the interaction region. The voltages required to achieve time focusing can be well approximated using Wiley-McLaren equations for a two-stage TOF mass spectrometer. Initially, the lens was kept at the drift voltage (lens OFF) to get temporal focusing. Next, the lens ring and drift voltages were optimized to achieve simultaneous time and space focusing. In the lens ON condition, we were able to get a complete $\text{4}\pi$ collection of singly charged ions up to 8 eV kinetic energy. However, this energy was restricted to 5 eV in the lens OFF condition. In figure \ref{Fig.3} we have shown the trajectories of singly charged nitrogen ions (N$^{\text{+}}$) having energies of 0.5 eV, 1.5 eV, and 3 eV ejected perpendicular to the spectrometer axis in the lens ON and OFF conditions. Although the lens voltage does not affect the time focusing condition, the change in ion-hit position on the detector can be easily seen (figure \ref{Fig.3}). The lens voltage also modifies the mean TOF of the ions. The change in TOF is a function of mass to charge ratio of the recoil ions. Figure \ref{Fig.4}(a) shows the experimentally measured TOF of N$^{\text{+}}$/N$_{\text{2}}^{\text{2+}}$ ion produced due to the fragmentation/ionization of N$_{\text{2}}$. The measured peak-to-peak difference in TOF of N$^{\text{+}}$/N$_{\text{2}}^{\text{2+}}$ ions is $\sim$ 50 ns. This is in excellent agreement with the time difference of 48 ns observed in simulations.


The initial 3D momenta ($p_x, p_y,$ and $p_z$) of the ions can be reconstructed from the TOF and detector hit positions using the following equations:

\begin{equation}
\begin{split}
    p_x &= \frac{mX}{T},\\
    p_y &= \frac{mY}{T},\\
    p_z &= -qE_s\Delta T,
\end{split}
\label{momentum}
\end{equation}
where $m$ and $q$ are the mass and charge of the recoil ion, $X$ and $Y$ are the hit positions of the recoil ion on the detector, and $T$ is the TOF. $E_s$ is the electric field in the extraction region. $\Delta T$ is the TOF difference $(T-T_0)$ between a fragment ion having a finite momentum $(p_z \neq 0)$ and the same ion having a zero initial momentum along the spectrometer axis $(p_z = 0)$. The equations are applicable for an ideal Wiley-McLaren type spectrometer. The modified equations considering the focusing effect of the lens ring in the drift tube region are given as:
\begin{equation}
    \begin{split}
    p_x &= \frac{mX}{fT_L}, \\
    p_y &= \frac{mY}{fT_L},\\
    p_z &= -CqE_s\Delta T_L,
\end{split}
\end{equation}
The quantities $f$ and $C$ are the correction factors arising due to the lens. These correction factors are determined from the simulation. The factor $f$ represents the magnification of the electrostatic lens in the detector plane. It is obtained by taking the ratio of ion-hit radii on the detector in lens ON and OFF conditions. Figure \ref{Fig.4}(b) is a plot of the radii in lens ON vs. lens OFF conditions for various initial energies of the recoil ions. The slope of this curve gives the value of $f$, which has been defined earlier as the "bending coefficient" \cite{lebech_2002, prumper_2007}. For our spectrometer the factor $f$ = 0.643. The factor $C$ comes from the modifications in the TOFs of the recoil ions due to the lens voltage. It is defined as the ratio of TOF with the lens OFF to that with the lens ON. For our spectrometer, the factor $C$ = 0.985. Bomme \emph{et al} \cite{bomme_2013} have pointed out that recoil ions with $p_z = 0$ but finite $p_x$ and $p_y$ will traverse different regions of the lens field, which may lead to the difference in TOF to be a function of the hit radius (R). Accordingly, the TOF difference is given by $\Delta T=T_{0L}(R)-T_{0L}(0)$, where $T_{0L}(R)$ is the TOF of recoil ions with $p_z = 0$ but finite $p_x$ and $p_y$, and $T_{0L}(0)$ is the TOF of recoil ions with $p_x=p_y=p_z = 0$. We performed detailed simulations to quantify the effect of finite momentum in the transverse direction of the spectrometer for several combinations of recoil ion mass and charge. The difference $\Delta T(=T_{0L}(R)-T_{0L}(0))$ is estimated to be $\le$ 4 ns. This corresponds to a momentum uncertainty of $\sim$ 5 a.u. which is much smaller than the resolution of the spectrometer, and thus can be neglected in momentum calculations.

\subsection{Data Acquisition}

\begin{figure}
\raggedright
\includegraphics[width=\textwidth]{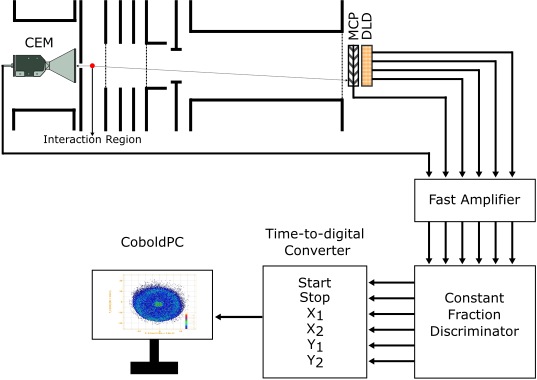}
\caption{Schematic diagram of the data acquisition system.}
\label{Fig.5}
\end{figure}

A block diagram of the data acquisition scheme is shown in figure \ref{Fig.5}. When a projectile ion interacts with a target atom or molecule, it ionizes the target producing electrons and recoil ion. The electrons are attracted toward the pusher electrode and are detected by the CEM. The recoil ions produced in the interaction region are pulled toward the MCP-DLD assembly by the electric field applied in the RIMS. The MCP produces a signal once it detects an ion and the DLD produces four signals $(x_1, x_2, y_1, y_2)$ corresponding to the ion-hit position. These six signals, one from the CEM, one from the MCP, and four from the DLD, are fed to a fast amplifier, and the amplified signals are given to constant fraction discriminators (CFDs). The outputs of the CFDs are sent to the TDC8HP module, which is an eight-channel multi-hit time-to-digital converter (TDC) \cite{roentdek}. The electron signal from the CEM acts as a "start" signal, whereas the MCP signal corresponding to the detection of an ion acts as a "stop" signal. The TDC8HP works in a multi-hit mode and can accept multiple "stops" for a single "start" signal. For multi-hit detection, each channel of the TDC is kept open for a 10 $\mu s$ window. During the experiment, the projectile ion beam current is kept low enough ($\sim$ 100 pA) such that the average electron and ion detector count rates remain below 5 kHz. This low beam-current results in an average coincidence rate of $\sim$ 300 - 500 Hz. The detection of an electron and the corresponding recoil ions produced in a molecular fragmentation is termed as an event. The ions detected within the TDC time window are considered to be originated from the same event. The TDC outputs corresponding to each event are recorded as a list mode file using the CoboldPC software \cite{roentdek}. The CoboldPC software is also used for further data processing and analysis.

\section{Results and Discussions }



\subsection{Spectrometer Mass Resolution}
\begin{figure*}
\raggedright
\includegraphics[width=\textwidth]{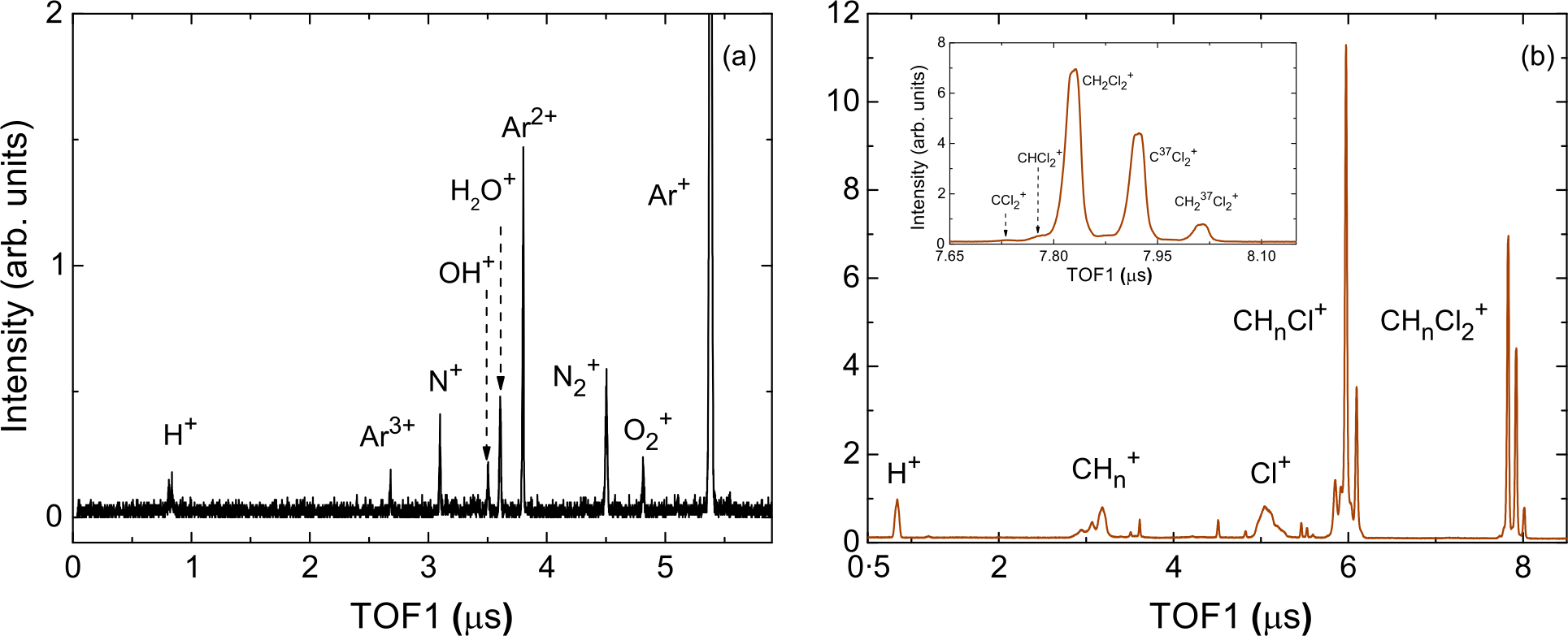}
\caption{First-hit TOF spectrum for 1 MeV proton impact on (a) argon (Ar) and (b) dichloro methane (CH$_2$Cl$_2$). The inset in (b) shows the magnified TOF spectrum for CH$_{\text{n}}$Cl$^{2+}_{2}$ (n=0,1,2).}
\label{Fig.6}
\end{figure*}

Figure \ref{Fig.6}(a) shows the first-hit TOF spectrum of atomic Argon (Ar) target under the impact of 1 MeV proton beam. Along with Argon ions (up to triply charged), several other peaks such as O$_{\text{2}}^{\text{+}}$, N$_{\text{2}}^{\text{+}}$, H$_{\text{2}}\text{O}^{\text{+}}$, OH$^{\text{+}}$, N$^{\text{+}}$, and H$^{\text{+}}$ can also be identified, which correspond to the background gases. The width (FWHM) of the Ar$^+$ ion TOF is used to estimate the mass resolution of the spectrometer. The FWHM $(\Delta T)$ of the TOF peak translates to a mass resolution $(\frac{M}{\Delta M}=\frac{T}{2\Delta T})$ of 198. In figure \ref{Fig.6}(b) we have also shown the TOF spectrum of a polyatomic target molecule, CH$_{\text{2}}$Cl$_{\text{2}}$, under identical experimental conditions and spectrometer potentials. The various isotopic contributions from $^{\text{37}}$Cl and $^{\text{13}}$C are clearly visible in the singly charged recoil ion peak of CH$_{\text{n}}$Cl$_{\text{2}}^+$ (n=0,1,2).

\subsection{Fragmentation of $\text{N}_{\text{2}}$}
\begin{figure*}
\raggedright
\includegraphics[width=\textwidth]{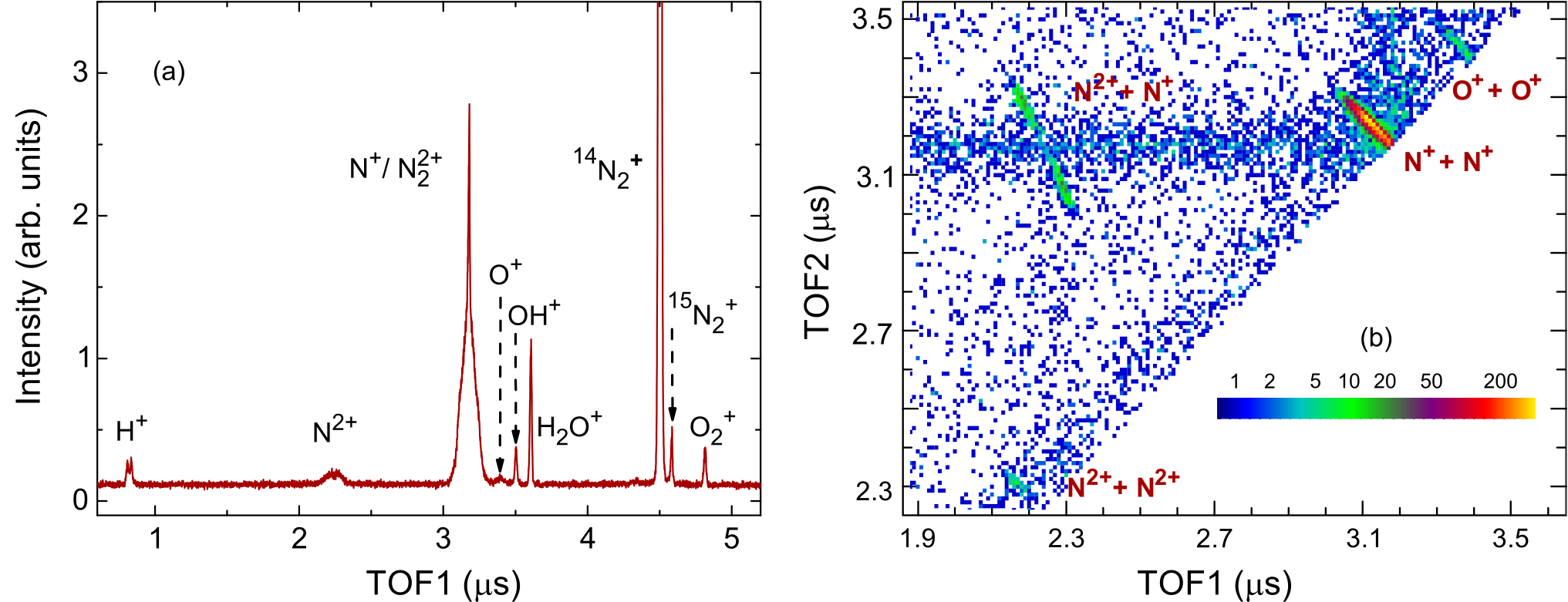}
\caption{(a) First hit TOF spectrum and (b) Ion-ion coincidence diagram for 1 MeV proton impact on molecular nitrogen (N$_{\text{2}}$).}
\label{Fig.7}
\end{figure*}

Ionization and subsequent fragmentation of homoatomic N$_{\text{2}}$ molecule has been studied extensively in the past. Hence, the N$_{\text{2}}$ molecular target serves as a benchmark system to evaluate and characterize the performance of a recoil ion momentum spectrometer. The first hit TOF spectrum of molecular N$_{\text{2}}$ under the impact of 1 MeV proton is shown in figure \ref{Fig.7}(a). Recoil ion peaks corresponding to multiple ionization and fragmentation products such as N$_{\text{2}}^{\text{+}}$, N$_{\text{2}}^{\text{2+}}$/N$^{\text{+}}$, N$^{\text{2+}}$ are visible. The TOF spectrum also has peaks corresponding to O$_{\text{2}}$ and water vapor present in the background. For a homoatomic molecule such as N$_{\text{2}}$, the recoil ions N$_{\text{2}}^{\text{2+}}$ and N$^{\text{+}}$ will have identical time-of-flight due to equal mass to charge ratio. However, the two recoil ions, N$_{\text{2}}^{\text{2+}}$ and N$^{\text{+}}$, will have different KER. The molecular ion N$_{\text{2}}^{\text{2+}}$ is created in direct collision with the energetic projectile and has negligible kinetic energy. Whereas the atomic ion N$^{\text{+}}$ is created predominantly via Coulomb fragmentation of N$_{\text{2}}^{\text{q+}}$ (q=2,3) molecular ions and carries large kinetic energy. This difference in KER of the two ions is visible in the shape of the peak corresponding to N$_{\text{2}}^{\text{2+}}$/N$^{\text{+}}$ in the TOF spectrum. The narrow peak at the center is due to the parent N$_{\text{2}}^{\text{2+}}$ molecular ion, which has small kinetic energy, and the broader structure is due to the energetic N$^{\text{+}}$ ions coming from different fragmentation channels.

In charged particle induced ionization of molecules, mainly three processes, pure ionization, pure capture, and transfer ionization, contribute to the total ionization yield. In the present experimental setup, the data acquisition system works in electron "start" mode. Therefore, the data is collected only for events corresponding to ionization and transfer ionization processes,  where at least one electron is ejected after the interaction between the projectile ion and the target molecule. Recoil ions created due to electron capture to the projectile are lost. Furthermore, the parent molecular ion may be created in an excited state depending on the energy transfer during the collision. If this state is repulsive in nature, then the molecular ion dissociates into fragments. This type of dissociation is known as the Coulomb explosion. In our experiment with 1 MeV proton on N$_{\text{2}}$, we have observed two fragmentation channels:
\begin{equation}
    {\text{N}}_{\text{2}}^{\text{2+}} \longrightarrow {\text{N}}^{\text{+}} + {\text{N}}^{\text{+}}
\label{ch1}
\end{equation}
\begin{equation}
    {\text{N}}_{\text{2}}^{\text{3+}} \longrightarrow {\text{N}}^{\text{2+}} + {\text{N}}^{\text{+}}
\label{ch2}
\end{equation}



As discussed previously, the TDC8HP module is capable of recording data corresponding to all fragment ions created in the same event. Various fragmentation channels following the ionization and subsequent fragmentation of the N$_{\text{2}}$ molecules can be distinguished in an ion-ion coincidence map or correlation diagram. For a diatomic molecule, each fragmentation event results in two hits on the detector. In the coincidence map, the TOF of the second hit (TOF2) is plotted against the TOF of the first hit (TOF1). On this 2d map, each point corresponds to a detection of two fragments that come from a single event of fragmentation of a molecular ion. Each fragmentation channel forms its own island on this map. Figure \ref{Fig.7}(b) shows the ion-ion coincidence map for the fragmentation of N$_{\text{2}}$. Here, we observe sharp traces for N$^{\text{+}}$+N$^{\text{+}}$ and N$^{\text{2+}}$+N$^{\text{+}}$ fragmentation channels. A less prominent trace of N$^{\text{2+}}$+N$^{\text{2+}}$ can also be identified (lower left quadrant in figure \ref{Fig.7}(b)). In addition, a small island of O$^{\text{+}}$+O$^{\text{+}}$ fragmentation channel is also visible. Traces corresponding to higher-order fragmentation channels have not been observed in the present work. The shape and slope of these traces depend on the underlying fragmentation dynamics of the precursor molecular ion. In the case of N$_{\text{2}}^{\text{q+}}$ the fragmenting atomic ions move in the opposite direction due to momentum conservation. The ratio of TOF of the two ions can be written as:
\begin{equation}
    Slope = \frac{\Delta T_2}{\Delta T_1}=\frac{p_{2z}q_{1}}{q_{2}p_{1z}}
\label{slope}
\end{equation}
Here, $\Delta T$ is defined as in equation (\ref{momentum}) and $p_{iz}$ is the momentum of the i$^{\text{th}}$ ion along the spectrometer axis. For Coulomb explosion (CE) $p_{1z}=-p_{2z}$ and hence the slope of the ion-ion coincidence plot is given as:

\begin{equation}
    Slope = - \frac{q_{1}}{q_{2}}.
\end{equation}  



For N$^{\text{+}}$+N$^{\text{+}}$ fragmentation channel, the slope of the coincidence trace is -1 and for N$^{\text{2+}}$+N$^{\text{+}}$, it is -2. This relationship is also used to filter out the false coincidences in the data.

\begin{figure*}
\raggedright
\includegraphics[width=\textwidth]{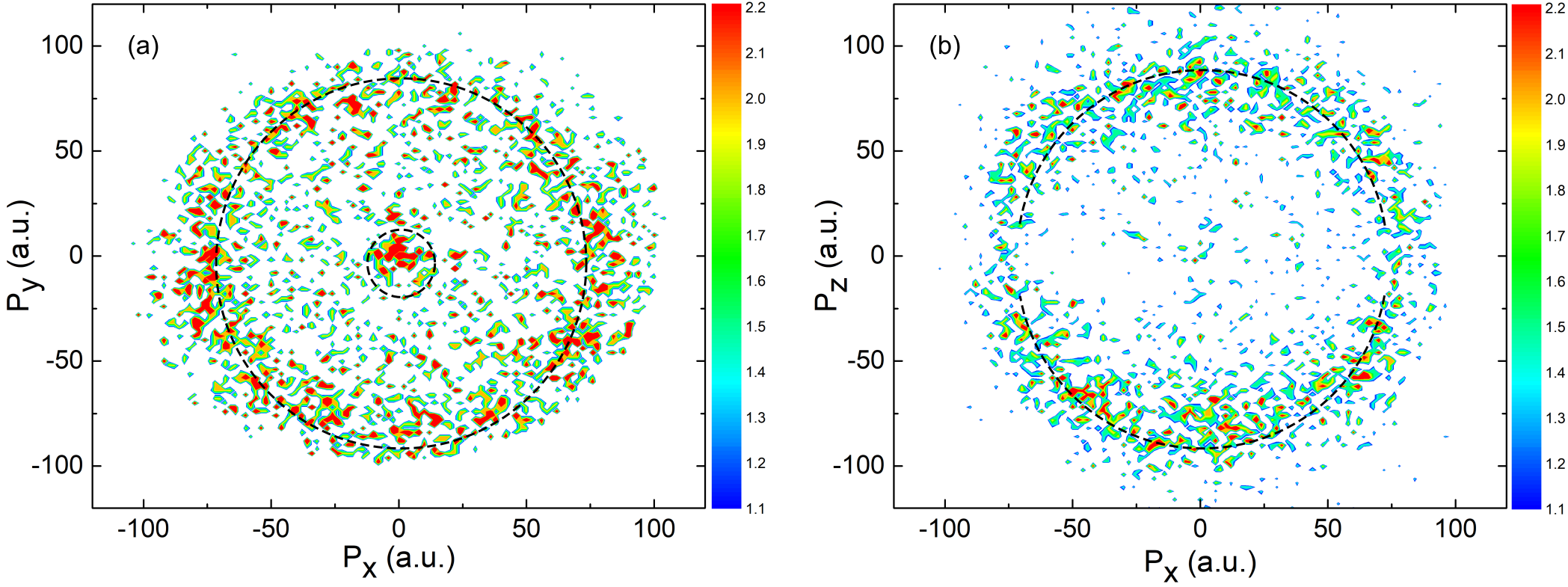}
\caption{(a) The P$_{\text{y}}$-P$_{\text{x}}$ and (b) P$_{\text{z}}$-P$_{\text{x}}$ momentum distribution of N$^{+}$ ions for 1 MeV proton impact on molecular nitrogen (N$_{\text{2}}$).}
\label{Fig.8}
\end{figure*}

The momentum distribution for the N$^{\text{+}}$+N$^{\text{+}}$ fragmentation channel is shown in figure \ref{Fig.8}. Fragment ion distribution in the plane perpendicular to the spectrometer axis (P$_{\text{y}}$ vs. P$_{\text{x}}$) shows isotropic behavior owing to the random orientation of the target gas molecules in the laboratory frame. The circular region at the center of the distribution (confined within the dashed line) corresponds to the recoil ions with very low kinetic energy. The outer circular ring originates from the recoil ions with higher kinetic energy. Similarly, one can also plot the momentum distribution in the P$_{\text{z}}$ - P$_{\text{x}}$ plane as shown in figure \ref{Fig.8}(b). The data with positive $p_z$ values are the ions with their initial momenta directed toward the detector along the spectrometer axis (the first hit), and those with negative $p_z$ values correspond to initial momenta directed away from the detector (the second hit) in the same fragmentation event. The plot also shows that the region around $p_z = 0$ has very few events. This region corresponds to the fragment ion pairs, which are ejected perpendicular to the spectrometer axis and arrive at the ion detector almost at the same time. However, due to the finite dead time of the MCP detector, the second hit is not recorded, resulting in the absence of data in this region. From the plot in figure \ref{Fig.8}(b), we estimate a detector dead time of $\sim$ 35 ns. 

\subsection{Kinetic Energy Release Distribution}
The 3D momenta of both the ions in a fragmentation event can be used to calculate the kinetic energy of the ions. as:
\begin{equation}
    KE_{i}=\frac{p^2_{ix}+p^2_{iy}+p^2_{iz}}{2m_{i}},
\end{equation}
\label{eq:KE}
where $p_{ix}$, $p_{iy}$, $p_{iz}$ are the 3D momenta of the recoil ions ($i$ = 1 for the first hit and $i$ = 2 for the second hit). The kinetic energy release (KER) for a particular fragmentation channel is the sum of the kinetic energies of the ions from the same event. \begin{equation}
    KER=\sum_{i} KE_{i}.
\end{equation}
\label{eq:KER}
We have analyzed the KER spectra of doubly and triply ionized nitrogen molecules.

\begin{figure*}
\raggedright
\includegraphics[width=\textwidth]{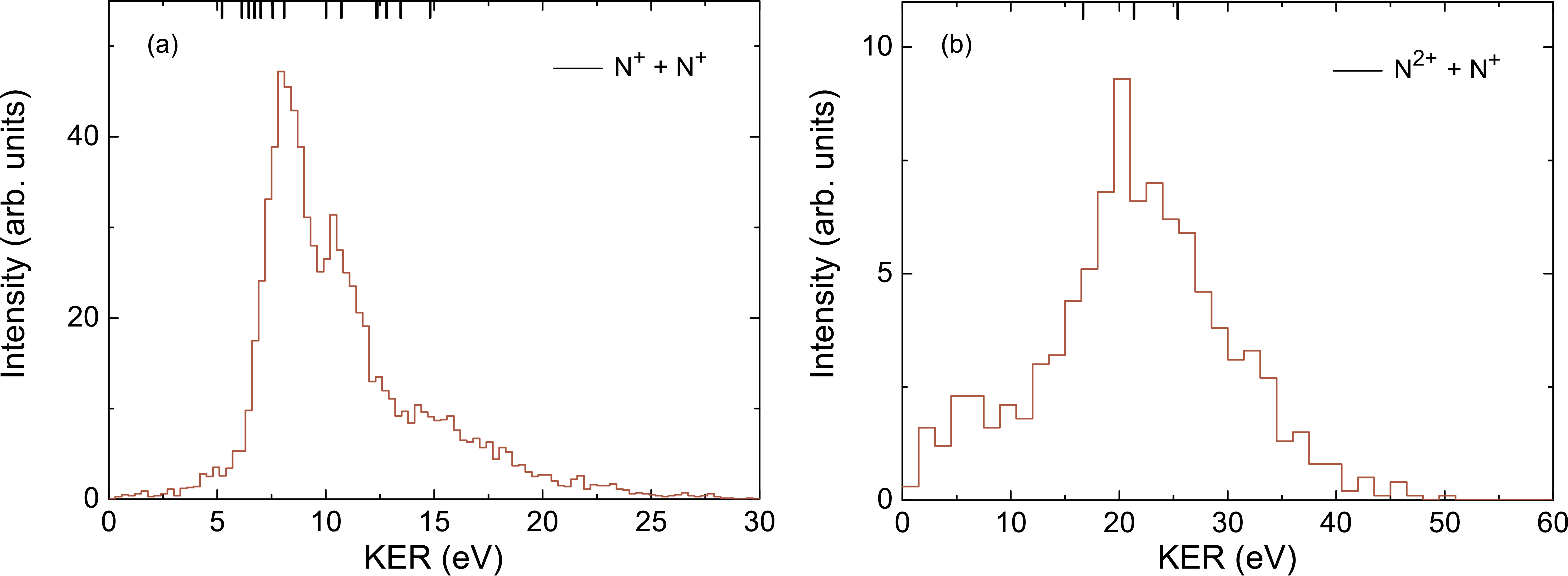}
\caption{Kinetic energy release distributions (KERDs) for the two fragmentation channels: (a) N$_{\text{2}}^{\text{2+}}$ $\rightarrow$ N$^{\text{+}}$+N$^{\text{+}}$ and (b) N$_{\text{2}}^{\text{3+}}$ $\rightarrow$ N$^{\text{2+}}$+N$^{\text{+}}$. The vertical lines represent calculated KER values \cite{pandey_2014, Rajput_2006,Safvan_1994}}
\label{Fig.9}
\end{figure*}

The kinetic energy release distribution (KERD) for the N$_{\text{2}}^{\text{2+}}$ $\rightarrow$ N$^{\text{+}}$+N$^{\text{+}}$ channel is shown in figure \ref{Fig.9}(a). The KERD shows distinct peaks around 7.8 eV, 10.5 eV, and 12.3 eV. The KER spectrum extends up to 20 eV on the higher energy side. The opserved KER spectrum is in very good agreement with the existing theoretical and experimental results \cite{lundqvist_1996,pandey_2014, Rajput_2006, khan_2015}. 
The peak around 7.8 eV can be attributed to the tunneling vibrational levels of several electronic states such as: 1$^\text{3}\Sigma_{\text{u}}^{+}$, 1$^{\text{3}}\Pi_{\text{g}}$, and 2$^{\text{1}}\Sigma_{\text{g}}^{+}$ all with the same dissociation limit of N$^{+}(^{\text{3}}\text{P})$ + N$^{+}(^{\text{3}}\text{P})$. On the other hand, the peak around 10.5 eV is attributed to the 3$^{\text{1}}\Sigma_{\text{g}}^{+}$ and 1$^{\text{1}}\Pi_{\text{g}}$ repulsive states. The 3$^{\text{1}}\Sigma_{\text{g}}^{+}$ state has a dissociation limit of N$^{+}(^{\text{1}}\text{D})$+N$^{+}(^{\text{1}}\text{D})$ producing a KER of 10.01 eV and the 1$^{\text{1}}\Pi_{\text{g}}$ state has a dissociation limit of N$^{+}(^{\text{3}}\text{P})$+N$^{+}(^{\text{3}}\text{P})$ producing a KER of 10.72 eV. The peak around 12.3 eV may have contributions from the 2$^{\text{3}}\Sigma_{\text{u}}^{+}$ and 2$^{\text{3}}\Pi_{\text{g}}$ states having a dissociation limit of N$^{+}(^{\text{3}}\text{P})$+N$^{+}(^{\text{3}}\text{P})$ and N$^{+}(^{\text{3}}\text{P})$+N$^{+}(^{\text{1}}\text{D})$, respectively. The KER values in the high energy region can be attributed to high lying repulsive states in the Frank-Condon region as calculated by Pandey \emph{et al} \cite{pandey_2014}. In table \ref{Table I} we have summarized our experimentally measured KER values for N$_{\text{2}}^{\text{2+}}$ and N$_{\text{2}}^{\text{3+}}$ fragmentation.


\begin{table}
\begin{tabular*}{\textwidth}{l  @{\extracolsep{\fill}}  c c r }
\hline\hline \\ 
Fragmentation & Most probable & Coulomb  & Relative  \\ 
channel       & KER           & energies & intensity \\ 
       	      & (eV)          & (eV)     & $(\%)$    \\ [0.05ex]\\
\hline \\ 

N$^{+}$+N$^{+}$ & 7.8 $\pm 0.3$    & 13.1 \footnote{\label{note1}At equilibrium N-N bond length of 1.098 \AA} & 87  \\ 
N$^{\text{2+}}$+N$^{+}$      & 19.5$\pm1$   & 26.2\textsuperscript{ a}    & 13  \\ [0.05ex]\\
\hline \\
Cl$^{+}$ + CH$_{2}$Cl$^{+}$  & 4.5 $\pm$ 0.4 &8.1 \footnote{\label{note1}At equilibrium C-Cl bond length of 1.77 \AA}  &93 \\
H$^{+}$+CHCl$_{\text{2}}^{+}$   & 4.5 $\pm$ 1.5 &13.2 \footnote{\label{note1}At equilibrium C-H bond length of 1.09 \AA}  &2    \\
$^\text{37}$Cl$^{+}$+CH$_\text{2}^\text{37}$Cl$^{+}$ &4.2 $\pm$ 0.4  &8.1 \textsuperscript{ b} &4\\
  [0.05ex] \\
	                                	
\hline \hline
	\end{tabular*}
	\caption{Observed most probable values in the KER spectra for the various fragmentation channels for N$_{\text{2}}^{\text{q+}}$ (q=2,3) and CH$_{\text{2}}{\text{Cl}}_{\text{2}}^{\text{2+}}$ induced by 1 MeV proton impact.The values calculated from a pure Coulomb explosion model are also included. The energy expected from this model is given by: E(eV)=14.4$\frac{{\text{q}}_{\text{A}}{\text{q}}_{\text{B}}}{{\text{R}}_{\text{e}}(\AA)}$, where q$_{\text{A}}$ and q$_{\text{B}}$ are the asymptotic charges of the two fragments, and R$_{\text{e}}$ is the equilibrium internal nuclear distance of the neutral molecule.}
	\label{Table I}
\end{table}

\subsection*{Angular Distribution of N$_{\text{2}}^{\text{2+}}$ $\rightarrow$  N$^{\text{+}}$+N$^{\text{+}}$}
The electron cloud in a molecule lacks spherical symmetry. As a consequence, the molecular ionization cross section depends on the orientation of the molecular axis (the internuclear axis for diatomic molecules) relative to the projectile ion beam direction. In collisions with fast projectiles, the molecular fragmentation following multiple ionization, takes place over the vibrational time scale ($\sim$ 10$^{\text{-14}}$ s) which is much faster compared to the rotational time period ($\sim$ 10$^{\text{-12}}$ s). Therefore the molecular axis remains frozen in space during ionization and subsequent fragmentation. As a result, the recoil ions produced due to fragmentation move along the molecular axis. Hence, for diatomic molecules, the measured 3D momentum vectors of the coincident ion pairs can be used to plot the angular distribution of the molecular ionization process. For the present collision system (E$_{\text{p}}$ = 1 MeV, v$_{\text{p}} \sim $ 6 a.u.), the Born approximation is valid and one can write the angular distribution as dipole type \cite{Siegmann_2002} as shown below:
\begin{equation}
    I(\theta) = I_0 [1 + \beta P_2(\cos \theta)]\sin \theta
    \label{ang_dist}
\end{equation}
Here, $I(\theta)$ is the ionization cross section at angle $\theta$, and $\beta$ is the angular asymmetry parameter. It has been shown earlier that P$_{\text{2}}(\cos \theta)$ is the only leading order contribution to the multipole expansion for 1 MeV projectile energy \cite{Edwards_1992}. The $\sin \theta$ factor in equation (\ref{ang_dist}) corresponds to the isotropic dependence of the ionization cross section on the molecular orientation. 
Several experimental and theoretical studies have reported anisotropy in the angular distribution in multiple ionization of diatomic molecules \cite{Dunn_1962, Werner_1997, Caraby_1997, Siegmann_2002}. However, the anisotropy has been mainly reported for triply and higher-order ionization of the diatom. For double ionization, the angular distribution shows a predominantly isotropic character. Siegmann \emph{et al} \cite{Siegmann_2003, Siegmann_2001, Siegmann_2002} have also shown that the anisotropy is inversely proportional to the projectile perturbation strength. 
For example, in collisions with 360 keV Xe$^{\text{18+}}$ on N$_{\text{2}}$ and O$_{\text{2}}$, they found no anisotropy, whereas in collisions with 770 MeV Xe$^{\text{18+}}$ significant orientation effects appeared at higher degrees of ionization. Recently Sharma \emph{et al} \cite{Sharma_2019} used 25 - 200 keV proton to study the fragmentation of multiply ionized CO$^{\text{n+}}({\text{n}} \leq {\text{4}})$ molecular ions. They found that the ionization cross section shows an anisotropy as well as asymmetry due to the heteroatomic nature of the CO molecule. As a result, ionization is favored when the ion beam first encounters the O atom.

\begin{figure}
\raggedright
\includegraphics[width=\textwidth]{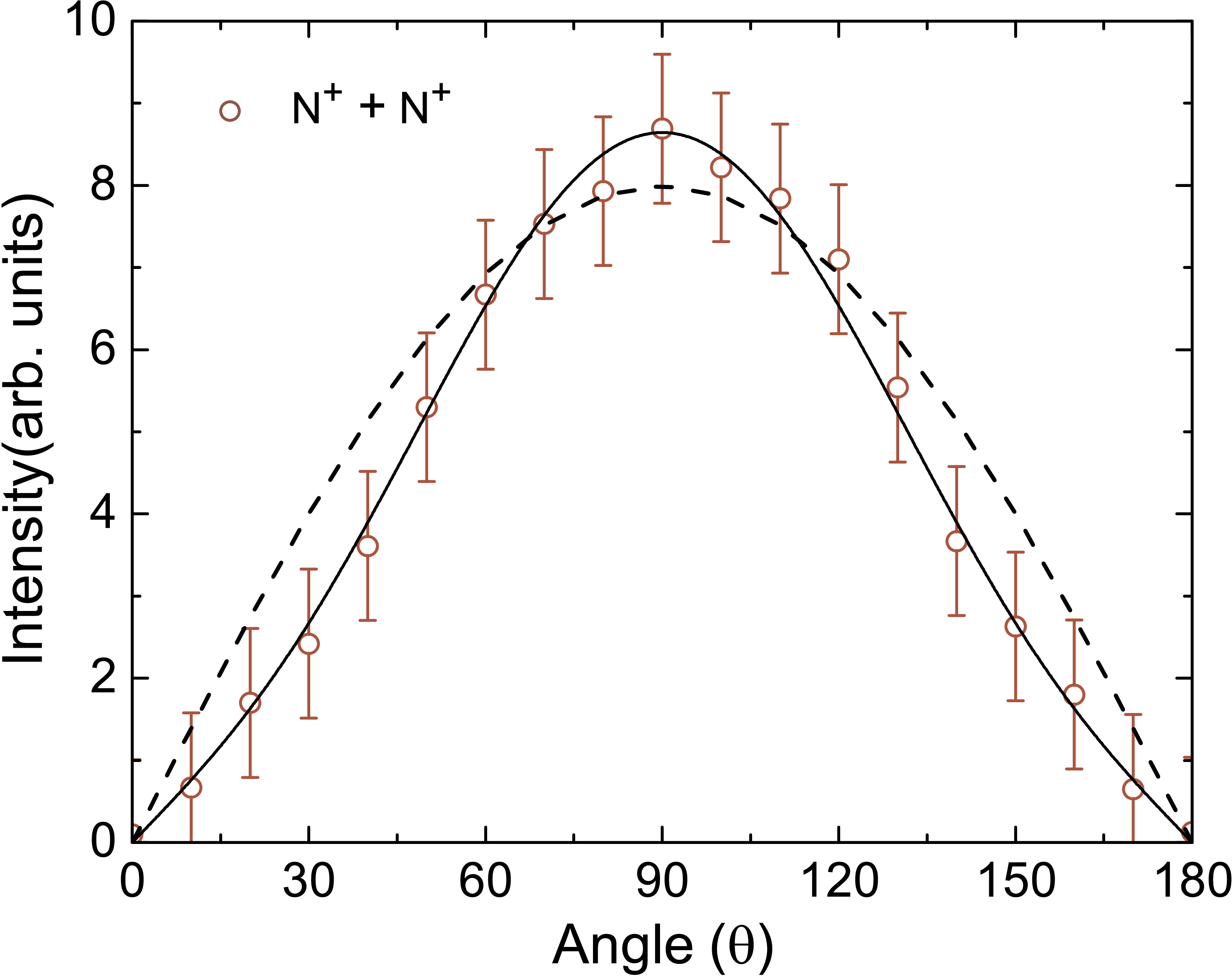}
\caption{Angular distribution for the N$_{\text{2}}^{\text{2+}}$ $\rightarrow$ N$^{\text{+}}$+N$^{\text{+}}$ fragmentation channel produced under the impact of 1 MeV protons. The open circles with error bars are the observed data, whereas the solid line is the fitted curve (see text). The dashed line shows the $sin \theta$ distribution representing the isotropic case.}  
\label{Fig.10}
\end{figure}

In figure \ref{Fig.10} we have shown the angular distribution for N$_{\text{2}}^{\text{2+}}$ production using momentum vectors for the N$^+$ + N$^+$ channel. The width of the total momentum distribution of the ion pair was used to estimate the angular resolution of the spectrometer to be $\sim {\text{10}}^{\circ}$. A curve corresponding to equation (\ref{ang_dist}) has been fitted to the measured angular distribution to estimate the value of asymmetry parameter $\beta$. Our estimated value of $\beta$ = -0.41 reveals the isotropic nature of the double ionization process with a slight preference for perpendicular orientation of the molecular axis. The value of $\beta$ is also slightly larger than that estimated by Siegmann \emph{et al} \cite{Siegmann_2002} for 300 keV proton impact.

\subsection*{KER distribution of N$_{\text{2}}^{\text{3+}}$ $\rightarrow$  N$^{\text{2+}}$+N$^{\text{+}}$}
The KERD for N$_{\text{2}}^{\text{3+}}$ decaying to N$^{\text{2+}}$+N$^{\text{+}}$ channel is shown in figure \ref{Fig.9}(b) (also see table \ref{Table I}). The KERD peaks around 20 eV and the KER values extend up to $\sim$ 50 eV. The spectrum also shows distinct features at $\sim$ 25 eV and $\sim$ 32 eV. There is also a low energy peak at $\sim$ 6 eV. Rajput \emph{et al} \cite{Rajput_2006} have calculated the potential energy curves (PECs) for N$_{\text{2}}^{\text{3+}}$ fragmentation channel. The peak at $\sim$ 20 eV is attributed to a few molecular states contributing to same final state configuration. The $^{\text{2}}\Sigma$ state of N$_{\text{2}}^{\text{3+}}$ can decay into N$^{\text{2+}}(^{\text{2}}{\text{P}})$+N$^{+}(^{\text{1}}{\text{S}})$ with a KER of 21.35 eV. In addition, the $^{\text{4}}\Sigma$ state can also produce a KER value of 19.77 eV by dissociating into N$^{\text{2+}}(^{\text{2}}{\text{P}})$+N$^{+}(^{\text{3}}{\text{P}})$. The $^{\text{2}}\Sigma$ state can also decay into N$^{\text{2+}}(^{\text{2}}{\text{P}})$+N$^{+}(^{\text{3}}{\text{P}})$ and produce a higher KER value of 25.40 eV. In an earlier calculation Safvan \emph{et al} \cite{Safvan_1994} have also shown that the $^{\text{2}}\Pi$ state of N$_{\text{2}}^{\text{3+}}$ can produce a KER of 16.65 eV with a dissociating limit of N$^{\text{2+}}(^{\text{1}}{\text{S}})$+N$^{+}(^{\text{3}}{\text{P}})$. The existing PECs for N$_{\text{2}}^{\text{3+}}$ have been calculated using only a few excited states of the molecular ion. However, contribution from high lying molecular states may lead to higher KER values.


\subsection{Fragmentation of CH$_{\text{2}}$Cl$_{\text{2}}$}




The capabilities of our spectrometer were further tested by investigating the ionization and fragmentation of a polyatomic molecule, CH$_{\text{2}}$Cl$_{\text{2}}$, in collisions with 1 MeV proton beam. The dichloro methane (CH$_{\text{2}}$Cl$_{\text{2}}$) molecule, in general, is used in many industrial applications such as food processing, refrigeration, aerosol formation, paint removers, etc. 
The molecule is a highly symmetric molecule having ${\text{C}}_{\text{2v}}$ symmetry. The two hydrogen atoms are located in a plane orthogonal to the plane formed by the two chlorine atoms and the carbon atom. The chlorine atom is more electronegative than hydrogen and carbon, and most of the electronic charge cloud is concentrated on the peripheral chlorine atoms \cite{vijayalakshmi_1997}. The ground electronic state of CH$_{\text{2}}$Cl$_{\text{2}}$ is X$^{\text{1}}{\text{A}}_{\text{1}}$. 

\begin{figure}
\raggedright
\includegraphics[width=0.9\textwidth]{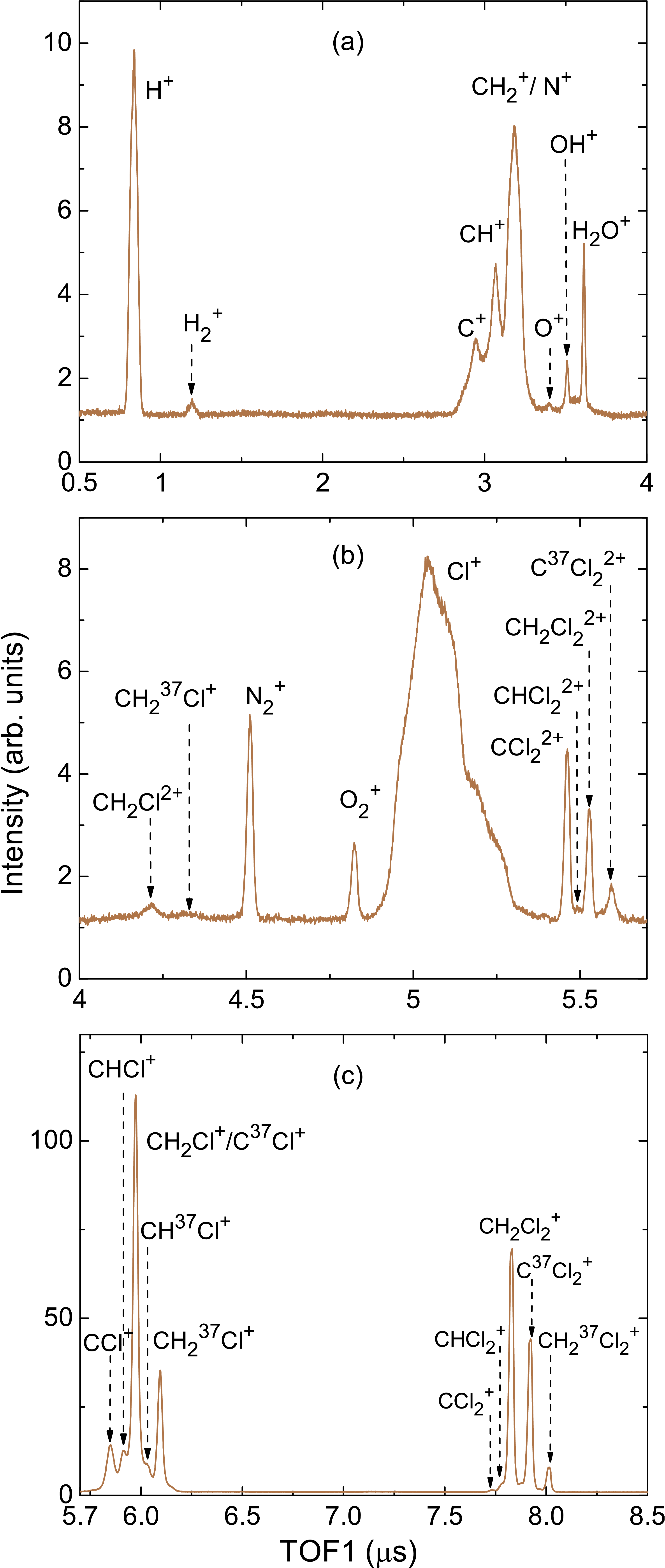}
\caption{First hit TOF specta for the ionization and fragmentation of dichloro methane (CH$_2$Cl$_2$) in collisions with 1 MeV protons. Each panel shows the expanded spectrum for different TOF values:(a) 0.5 - 4.0 $\mu s$, (a) 4.0 - 5.7 $\mu s$, and (a) 5.7 - 8.5 $\mu s$.}
\label{Fig.11}
\end{figure}

The TOF spectra, shown in figure \ref{Fig.11}(a) - (c), is very rich with several closely lying peaks due to the polyatomic nature of the target molecule. Each panel shows an expanded region of the complete TOF spectrum (see figure \ref{Fig.6}(b)). The most prominent peaks in the TOF spectrum correspond to the CH$_{\text{n}}$Cl$^+$ molecular ion followed by the parent molecular ion CH$_{\text{n}}$Cl$_{\text{2}}^+$. Finer features resulting from H-atom loss and isotopic contributions of $^{\text{37}}$Cl accompany the two peaks. The CH$_{\text{n}}$Cl$^+$ peak has contributions from single ionization followed by neutral Cl-atom emission as well as Coulomb fragmentation of the doubly charged parent molecular ion CH$_{\text{n}}$Cl$_{\text{2}}^{\text{2+}}$, which is also evident from a substantial yield of Cl$^+$ atomic ion in the TOF spectrum. The lower mass region of the TOF spectrum is dominated by H$^+$ and CH$_{\text{n}}^+$ fragment ions. We also observe very small yields of H$_{\text{2}}^+$ molecular ion as well as intact CH$_{\text{2}}$Cl$_{\text{2}}^{\text{2+}}$ ion. Higher-order ionization and fragmentation channels have not been observed for the present collision system. Our measured TOF spectrum compares well with the TOF spectrum reported earlier by Bhardwaj \emph{et al} \cite{bhardwaj_1998} for 50 MeV Si$^{\text{3+}}$ projectile. In another experiment, Alcantara \emph{et al} \cite{alcantara_2011} had measured the ionization yields of various fragment ions for 200 keV - 2 MeV proton impact ionization of CH$_{\text{2}}$Cl$_{\text{2}}$. In table \ref{Table I} we have presented the relative yields of the fragment ions. The measured yield is in good agreement with those reported earlier \cite{alcantara_2011} for the same collision system.      

\begin{figure*}
\raggedright
\includegraphics[width=\textwidth]{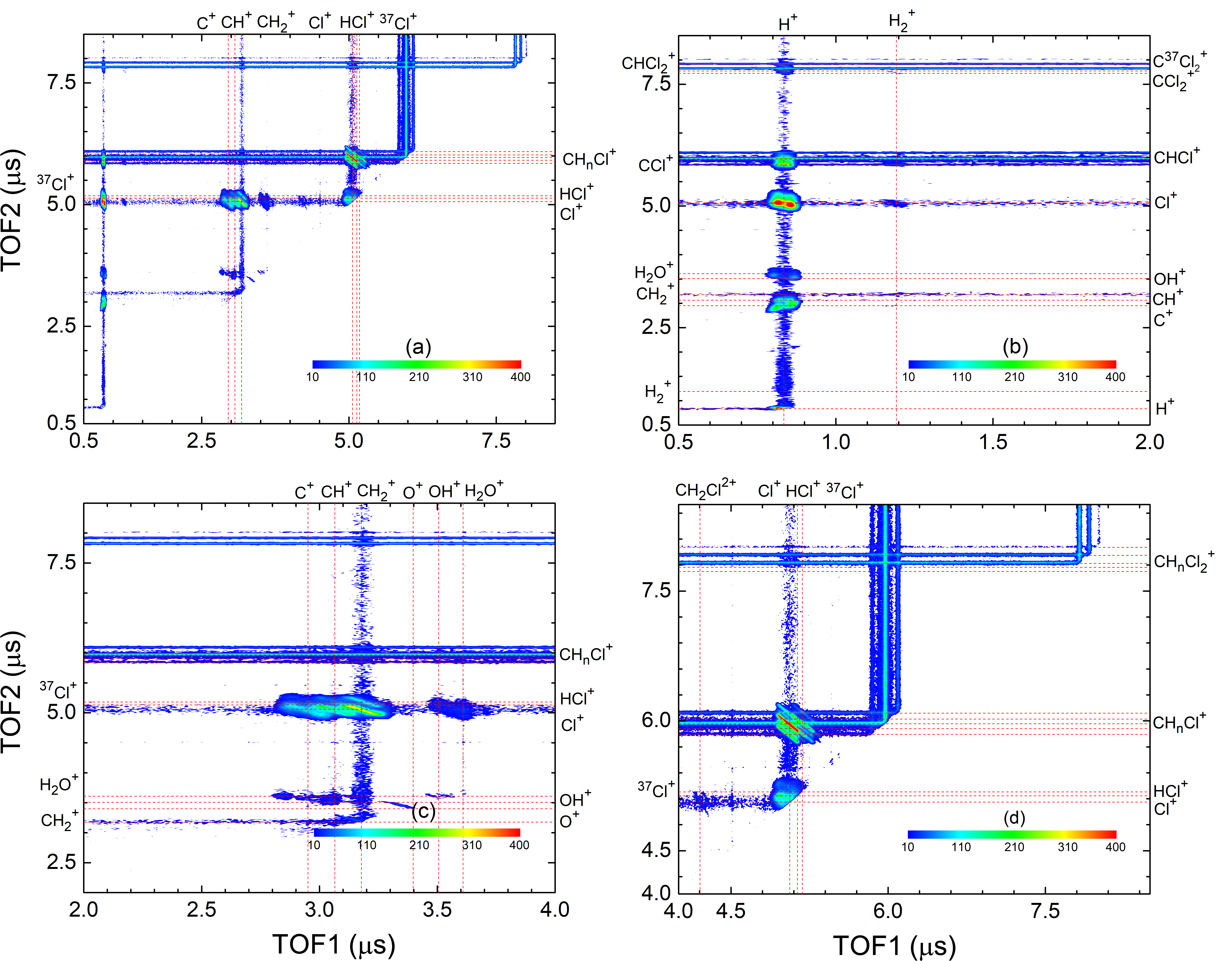}
\caption{Ion-ion coincidence diagram for the fragmentation of dichloro methane (CH$_2$Cl$_2$) in collisions with 1 MeV protons. Each panel shows this diagram for different TOF1 (first hit TOF) values: (a) 0.5 - 8.5 $\mu s$, (b) 0.5 - 4.0 $\mu s$, (c) 4.0 - 5.7 $\mu s$, and (d) 5.7 - 8.5 $\mu s$. }
\label{Fig.12}
\end{figure*}

The fragmentation process can be studied in more detail using the multi-hit coincidence plots. In figure \ref{Fig.12} we have shown the TOF coincidence plots corresponding to fragmentation of the CH$_{\text{n}}$Cl$^{\text{q+}}$ (${\text{q}} \ge {\text{2}}$) molecular ions. The coincidence plot shows several islands revealing various possible fragmentation channels following precursor ionization. The slopes and shapes of these islands carry the signature of the underlying fragmentation process. In figure \ref{Fig.12} panel (b) - (d), we have shown different regions of the coincidence plot for prominent fragmentation channels. As shown in panel (b), the H$^+$ ion appears in coincidence with CH$_{\text{n}}$Cl$^+$ ion, Cl$^+$ ion, and CH$_{\text{n}}^+$ ion. It is rather surprising that H$^+$ coincidence with CH$_{\text{n}}$Cl$_{\text{2}}^+$ molecular ion is feeble in the coincidence plot (see figure \ref{Fig.12} panel (a)). Compared to other coincidences, the extremely low yield of this channel suggests that Coulomb fragmentation of the doubly ionized parent molecular ion to H$^+$ with other recoil ions is accompanied by loss of neutral Cl atom. The islands corresponding to H$^+$ coincidence with CH$_{\text{n}}^+$ and H$^+$ with Cl$^+$ can not be due to a pure two-body breakup. The coincidences most likely originate from triple ionization and subsequent fragmentation of the parent molecular ion. More information about the processes can be obtained by considering the third fragment ion. However, the present analysis is restricted only up to two coincident fragment ions. In panel (c), one can identify a single prominent island group corresponding to the Cl$^+$ + CH$_{\text{n}}^+$ coincidence. The fragmentation channel may have a contribution from two body breakups following neutral Cl-atom loss upon precursor ionization. The slopes of the islands are in agreement with that expected from a pure two-body breakup (see equation (\ref{slope})). There is also a possibility of a three-body fragmentation following triple ionization for this coincidence channel. A detailed analysis of the momentum correlation of these fragment ions can reveal the nature of the process in terms of sequential or concerted fragmentation.

Panel (d) shows the region of pure two-body coincidences corresponding to Cl$^+$/HCl$^+$ + CH$_{\text{n}}$Cl$^+$ ion pairs. Here also, the slope of the coincidence plot matches with equation (\ref{slope}) in agreement with the two-body nature of the breakup. An interesting feature observed in the coincidence spectrum is the detection of HCl$^+$ fragment ion. In the parent molecule, there is no direct bonding between the Cl and H atoms. Therefore, HCl$^+$ production suggests ultra fast proton migration during the dissociation of CH$_{\text{2}}$Cl$_{\text{2}}^{\text{2+}}$. This process is possible in highly excited molecular ions where the bond angles deform significantly from their equilibrium geometry. Such ultra fast proton migration in heavy-ion impact ionization and fragmentation of methanol (CH$_{\text{3}}$OH) has been reported earlier by De \emph{et al} \cite{sankar_2006}. We have also observed the formation of H$_{\text{2}}^+$ molecular ion along with CCl$^+$ recoil ion fragment in the TOF coincidence plot. 

\subsection*{KER distribution for two-body dissociation of CH$_{\text{2}}{\text{Cl}}_{\text{2}}^{\text{2+}}$}

\begin{figure*}
\raggedright
\includegraphics[width=\textwidth]{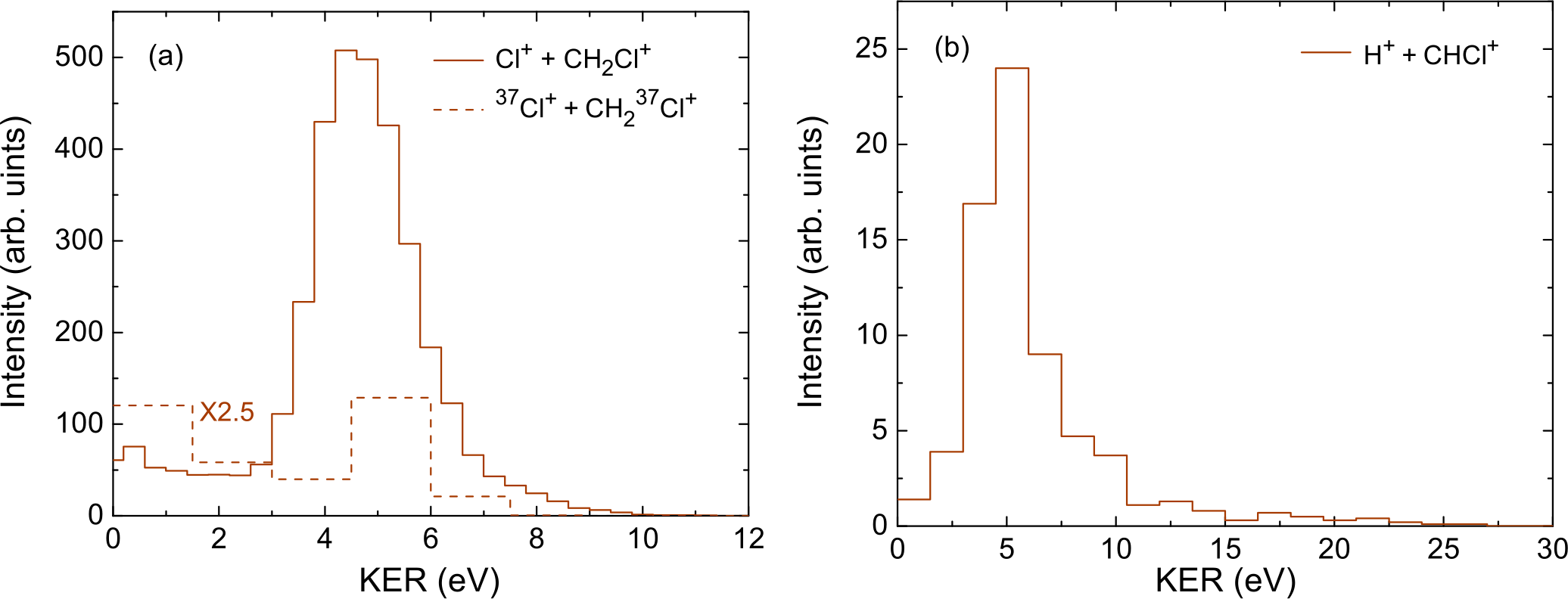}
\caption{Kinetic energy release distributions (KERDs) for the fragmentation channels: a) CH$_{\text{2}}$Cl$_{\text{2}}^{\text{2+}}$ $\rightarrow$ Cl$^+$ + CH$_{\text{n}}$Cl$^+$ and b) CH$_{\text{2}}$Cl$_{\text{2}}^{\text{2+}}$ $\rightarrow$ CH$_{\text{2}}$Cl$^{\text{2+}}$ + Cl $\rightarrow$ H$^+$ + CHCl$^+$. The dashed line in (a) shows the KERD for the isotopic fraction of CH$_{\text{2}}^{\text{37}}$Cl$_{\text{2}}^{\text{2+}}$.}
\label{Fig.13}
\end{figure*}

Figure \ref{Fig.13} shows the KERD for the two most prominent fragmentation channels as seen in the coincidence spectrum, i) CH$_{\text{2}}$Cl$_{\text{2}}^{\text{2+}}$ $\rightarrow$ Cl$^+$ + CH$_{\text{n}}$Cl$^+$ and ii) CH$_{\text{2}}$Cl$_{\text{2}}^{\text{2+}}$ $\rightarrow$ CH$_{\text{2}}$Cl$^{\text{2+}}$+Cl $\rightarrow$ H$^+$+CHCl$^+$. The KERD for channel (i) shows a most probable KER value of around 5 eV, along with a small peak structure around 1 eV. The C-Cl bond distance in the equilibrium geometry of CH$_{\text{2}}$Cl$_{\text{2}}$ is 1.77 \AA \hspace{0.5em} \cite{grant_1999,myers_1952}. Thus the most probable KER value according to the Coulomb explosion model is $\sim$ 8.1 eV. Similarly, figure \ref{Fig.13}(b) shows the KERD for channel (ii). This channel has a broad spread compared to channel (a), extending from 0 to about 25 eV. This broader range could be due to the high energy gained by the H$^+$ ion. The most probable kER for this channel is $\sim$ 5 eV.

\section{CONCLUSION}
We have presented the design and performance characteristics of a recoil ion momentum spectrometer suitable for studying molecular fragmentation following ion impact ionization. The spectrometer has been installed at the 1.7 MV Tandetron Accelerator Laboratory at IIT Kanpur. Detailed simulations were performed to understand the working parameters of the spectrometer. We have presented the fragmentation dynamics of a diatomic molecule N$_2$ to validate the performance of the spectrometer. Ion impact studies on the polyatomic CH$_2$Cl$_2$ molecule were also performed. Momenta and kinetic energy release distributions were derived for various fragmentation channels. The reported data is in good agreement with existing investigations.

\section{ACKNOWLEDGEMENT}
AHK would like to acknowledge the financial support received from Science and Engineering Research Board (SERB), Govt. of India via grant No. ECR/2017/002055. AD is thankful to Dr. Achim Czasch for his help in understanding the CoboldPC software. AHK would like to thank Dr. C. P. Safvan and Prof. L. C. Tribedi for their help. We also thank Mr. Sahan Sykam for the maintenance of the accelerator facility during the experiments.

\section{Author Contribution}
AHK and AD designed the spectrometer and carried out the simulations. AD, RT, and SB performed the experiments. AD and AHK analyzed the data and prepared the manuscript.

\bibliographystyle{utphys}
\bibliography{References}

\end{document}